\documentclass[twocolumn]{aastex631}
\usepackage{url}

\usepackage{multirow}  
\usepackage{graphicx}  
\usepackage{xspace}
\usepackage{natbib}
\usepackage{appendix}

\usepackage{amsmath}  
\usepackage{newtxtext,newtxmath}
\usepackage{bm,ctable,verbatim}
\usepackage{xcolor}
\usepackage{threeparttable}

\begin{document}

\title{Active Galactic Nucleus Feedback in an Elliptical Galaxy. IV. The Importance of the Jet-Wind Coupling}

\author[0009-0009-7849-2643]{\textsc{Minhang Guo}}
\affiliation{Shanghai Astronomical Observatory, Chinese Academy of Sciences, Shanghai  200030, People's Republic of China.}
\affiliation{ShanghaiTech University, 393 Middle Huaxia Road, Pudong, Shanghai 201210, People's Republic of China.}
\affiliation{University of Chinese Academy of Sciences, 19A Yuquan Road, Beijing 100049, People's Republic of China.}

\author[0000-0003-3564-6437]{Feng Yuan}
\affiliation{Center for Astronomy and Astrophysics and Department of Physics, Fudan University, Shanghai 200438, P.R.China}

\author[0000-0001-9658-0588]{Suoqing Ji}
\affiliation{Center for Astronomy and Astrophysics and Department of Physics, Fudan University, Shanghai 200438, P.R.China}
\affiliation{Key Laboratory of Nuclear Physics and Ion-Beam Application (MOE), Fudan University, Shanghai 200433, P.R.China}

\author[0000-0003-0900-4481]{Bocheng Zhu}
\affiliation{National Astronomical Observatories, Chinese Academy of Sciences, 20A Datun Road, Beijing 100101, P.R.China}

\correspondingauthor{Feng Yuan, Suoqing Ji}
\email{fyuan@fudan.edu.cn, sqji@fudan.edu.cn}
\newcommand \mdot{\dot M} 
\newcommand \rg{r_{g}}

\begin{abstract}
This is the fourth paper of our series investigating the effects of active galactic nucleus (AGN) feedback in the evolution of an elliptical galaxy using the {\it MACER} framework. While previous works considered only AGN radiation and wind, we now add jet feedback. The values of the jet parameters are taken from small-scale general relativity MHD simulations of black hole accretion. We run three models: {\tt FullFeedback}, {\tt JetOnly}, and {\tt WindOnly}. Time-averaged star formation rates are $10^{-1}$, $10^{-2}$, and  $10^{-3} \mathrm{M}_\odot\,\mathrm{yr}^{-1}$ in {\tt JetOnly}, {\tt WindOnly}, and {\tt FullFeedback}, respectively. Despite the higher jet power, jet feedback is less efficient than wind due to a small opening angle and low momentum flux. The much lower star formation rate in {\tt FullFeedback} indicates nonlinear coupling between jet and wind, with stronger suppression than the linear sum. The AGN energy dissipation efficiency values (fraction of injected kinetic energy dissipated via turbulence and shock) are 0.64 ({\tt FullFeedback}), 0.48 ({\tt WindOnly}), and 0.26 ({\tt JetOnly}). In the {\tt FullFeedback} model the wind-jet shear results in Kelvin-Helmholtz instability, driving stronger turbulence that effectively converts AGN kinetic energy into heating.
\end{abstract}


\section{Introduction}
\label{sec:intro}

The interplay between the central black hole and its surrounding gas in an elliptical galaxy, i.e., active galactic nucleus (AGN) feedback, constitutes a fundamental aspect of galaxy evolution, with feedback processes extending from the innermost regions to the galaxy outskirts \citep{Fabian1994, Peterson2006, 2012Fabian,2010Ostriker,2012Gaspari,2015Li,2015Prasad,2017Ciotti,2019Wang}.

This paper is the fourth in our series of works investigating AGN feedback in elliptical galaxies based on the MACER numerical simulation model \citep{2018arXiv180705488Y}. The first work is presented in \citet{2018Yuan}, which proposed the key physics of the MACER model. The model is developed from many early works \citep[e.g.,][]{2001Ciotti, 2009Ciotti, 2011Novak,2014Gan}. There are two key features of the model. One is that the inner boundary of the simulation domain of MACER is smaller than the Bondi radius. In this case, we can calculate the mass flux at the inner boundary. Combined with the theory of black hole accretion, we can reliably determine the black hole accretion rate, which determines the power of the central AGN and is the most important parameter in AGN feedback studies. The second key feature of MACER is that it incorporates the state-of-the-art AGN physics, including radiation and wind as a function of accretion rates in both the hot and cold accretion (feedback) modes. The effects of AGN feedback on the black hole growth, AGN light curve, and star formation are calculated and compared to observations. \citet{2018Yuan} deal with elliptical galaxies with a small angular momentum. In the second paper of this series, \citet{2018ApJ...864....6Y} extended \citet{2018Yuan} to the case of elliptical galaxies with a large angular momentum. These two papers do not include cosmological inflow. The effects of cosmological inflow were considered in the third paper --- \citet{2023Zhu}. They found that the inflow gas hardly enters the galaxy but is blocked beyond $\sim$20 kpc from the central galaxy and becomes part of the circumgalactic medium. In these three works, the jet is neglected. However, jets may carry significant power in the hot accretion mode where the AGN in elliptical galaxies mostly reside. As the fourth paper of this series, the aim of the present paper is to include jet in our MACER simulation and investigate its feedback effects. 

In fact, many numerical simulation works have explored the role of AGN jet in galaxy evolution. Nonprecessing jet simulations \citep{2004Omma, 2007Cattaneo, 2011Gaspari, 2017Choi, 2017Weinberger} show less effective energy dissipation in the surrounding medium. Later studies incorporated precessing jets \citep{2012Gaspari, 2014Lia, 2014Lib, 2016Yanga, 2016Yangb, 2017Bourne, 2023Husko}, wide-angle jets \citep{2015Prasad, 2017Hillel, 2018Hillel}, or spherical energy injections \citep{2015Reynolds} to enhance heating deposition. One main problem with these works is that the jet parameters adopted in the model are mostly free and lack a solid physical base. On the other hand, intensive studies of jet formation, especially 3D general relativity MHD (GRMHD) numerical simulations of jet formation have been performed in the past decades (\citealt[e.g.,][]{2010Tchekhovskoy,2016Monika, 2021Yang, 2021Liska,2019Chatterjee, 2024Yang}; see \citealt{2014Yuan} and \citealt{2020ARA&A..58..407D} for reviews). These works have obtained strict constraints on the properties of jets, including the mass flux, velocity, and opening angle. These constraints have not been taken into account in previous jet feedback simulations. Based on data of GRMHD simulations of black hole accretion, using a ``virtual test particle trajectory'' approach, \citet{2021Yang} have calculated their properties, which will be adopted in our present work.

In addition to the problem of jet parameters, another issue with previous works is that they usually have not taken into account jet and wind simultaneously. GRMHD simulations of hot accretion flows have revealed that most of the accreting gas does not fall onto the black hole horizon but is lost in wind \citep{2012ApJ...761..130Y, 2012Narayan, 2015Yuan, 2016Bu, 2021Yang}. These theoretical predictions have been confirmed by more and more observations \citep[e.g.,][]{2013Wang, 2016Cheung,2021Shi,2022ApJ...926..209S}. \citep{2021Yang} have compared the properties of wind and jet and found that, even in the case of a magnetically arrested disk (MAD, \citealt{2003Narayan}) around a very rapidly spinning black hole where the jet is supposed to be the strongest, the momentum flux of the jet is smaller than that of the wind and the total energy flux of the jet is larger than that of the wind by at most a factor of 10. This result indicates the importance of including wind in the feedback simulations. In addition to the important momentum and energy feedback, wind has also been invoked to confine and collimate jets \citep{2009Lyubarsky, 2016Globus,2019Park, 2021Chen}. Thus, the inclusion of wind is also necessary to obtain the correct dynamical state of the jet in the simulations. Moreover, in the present work, we find another reason why it is important to consider jet and wind simultaneously. That is, the coupling of jet and wind can result in strong turbulence, whose dissipation efficiently deposits their kinetic energy into the interstellar medium (ISM) in the galaxy. This process exhibits a conversion efficiency that is higher than the sum of their individual effects. 

In the present paper, we incorporate jet in the low-accretion hot mode into the MACER model, with its parameters taken from GRMHD simulations of black hole accretion, alongside the wind feedback already present in our model.
This paper is organized as follows. In Section \ref{sec:methods}, we describe the numerical setup and physical models, especially the jet model incorporated into the model. The main results are presented and analyzed in Section \ref{sec:results}, with particular focus on the jet-wind interaction and its effects on galaxy evolution. We discuss and summarize our conclusions in Section \ref{sec:diss}.

\section{Methods}
\label{sec:methods}

\subsection{Basic setup}

The basic setup of the simulation is similar to that of \cite{2018Yuan}, so we only briefly summarize it as follows. Our simulations focus on the evolution of isolated elliptical galaxies with a stellar mass $M_\star = 3 \ \times 10^{11} M_\odot$ and a supermassive black hole (SMBH) of $M_\mathrm{BH} = 4.5\ \times 10^9 M_\odot$ located in the galactic center. The galaxy model is an isolated elliptical galaxy with a spherically symmetric background gravitational potential, consisting of stellar components embedded in a dark matter halo. The stellar distribution is described by a Jaffe profile (\citealt{1983Jaffe}). 

The simulations are carried out in 2D axisymmetric spherical coordinates ($r$, $\theta$, $\phi$), where the $\phi$ coordinate is not an active degree of freedom — this is a ``2.5D'' setup where azimuthal components (e.g., $v_\phi$) can be present but there is no $\phi$ dependence. The radial grid spacing is logarithmically increasing, reaching a resolution of $\sim 0.3\,\mathrm{pc}$ at the inner boundary, and covering a radial range from $2.5\,\mathrm{pc}$ to $500\,\mathrm{kpc}$. The Bondi radius \cite{1952Bondi}, which is the outer boundary of accretion flow, is well within the simulation domain and is resolved by the highest resolution of $0.3\,\mathrm{pc}$ for almost all of the time. Such a setting has two advantages. One is that the accretion rate at the black hole horizon can be precisely calculated by combining the calculated mass inflow rate at the inner boundary and the theory of black hole accretion. This is crucially important since the black hole accretion rate determines the strength of the AGN and thus the power of feedback. The second advantage is that we can safely inject AGN outputs, namely, wind, jet, and radiation, at the inner boundary and calculate their interaction with the ISM of the galaxy, avoiding the use of some parameterized approach when calculating the interaction.  

The code ZEUS-MP \citep{2006Hayes} is used to solve the following governing equations: 
\begin{align}
  \frac{\partial \rho}{\partial t} + \nabla \cdot \left( \rho \bm{v} \right) &= \alpha_\star \rho_{\star} + \dot{\rho}_\mathrm{II}-\dot{\rho}_{\star}^{+}, \label{eq:mom} \\
  \frac{\partial \bm{p}}{\partial t} + \nabla \cdot \left( m\bm{v} \right) &= -\nabla p_{\rm gas} + \rho\bm{g} -\nabla p_{\rm rad} - \dot{\bm{m}}_{\star}^{+} \label{eq:energy} \\
  \frac{\partial E}{\partial t} + \nabla \cdot \left( E \bm{v} \right) &= -p_{\rm gas}\nabla \cdot \bm{v} + H - C + \dot{E}_\mathrm{I} + \dot{E}_\mathrm{II} + \dot{E}_{S} - \dot{E}_{\star}^{+},
\end{align}
where $\rho$, $\bm{v}$, $\bm{p}$, and $E$ are the gas density, velocity, momentum and internal energy, respectively. The gas pressure is determined by the ideal gas equation of state $p_{\rm gas} = \left( \gamma - 1 \right) E$, where the specific heat ratio is $\gamma = 5/3$, and $\bm{g}$ is the gravitational field of the galaxy (i.e., stars, dark matter, plus the contribution of the  central SMBH). Source terms are included in the equations: $\alpha_{\star} \rho_{\star}$ is the mass source from stellar evolution, $\dot{\rho}_\mathrm{II}$ is the recycled gas from supernovae (SNe II), and $\dot{E}_\mathrm{I}$ and $\dot{E}_{II}$ are feedback from SNe Ia and SNe II, respectively \citep{1989Mathews,2005Tang,2012Ciotti,2012Novak}.  The $H$ and $C$ terms denote heating and radiative cooling rates, respectively. The heating/cooling rate calculation follows \citet{2005MNRAS.358..168S} and \citet{2017Xie}, who describe the net heating and cooling per unit volume of the gas in photoionization equilibrium, including Compton heating/cooling, bremsstrahlung cooling, photoionization, and line and recombination cooling. \citet{2005MNRAS.358..168S} and \citet{2017Xie} deal with the AGN radiative heating via the calculation of a ``Compton temperature''  in the case of luminous and low-luminosity AGNs, respectively. The $\dot{\bm{m}}_{\star}^{+}$ and $\dot{E}_{\star}^{+}$ terms represent the mass and energy feedback from stellar evolution, respectively, where the age of the stellar population is $2\,\mathrm{Gyr}$, and the star formation criteria and rates are calculated according to previous studies \citep{2011Novak, 1998ApJ...498..541K,2007ApJ...654..304K} (see \citet{2019ARA&A..57..227K} for a review). We refer to \citet{2018Yuan} for more details on other aspects of the model, including the AGN physics, calculation of star formation rate (SFR), SN feedback, and stellar evolution/feedback. Cosmological inflow is not included in this work since it usually does not have a strong effect on the SFR \citep{2023Zhu}. 

\subsection{Updated AGN physics in {\it MACER}: the inclusion of a jet}

\begin{figure*}
    \centering
    \includegraphics[width=0.95\textwidth]{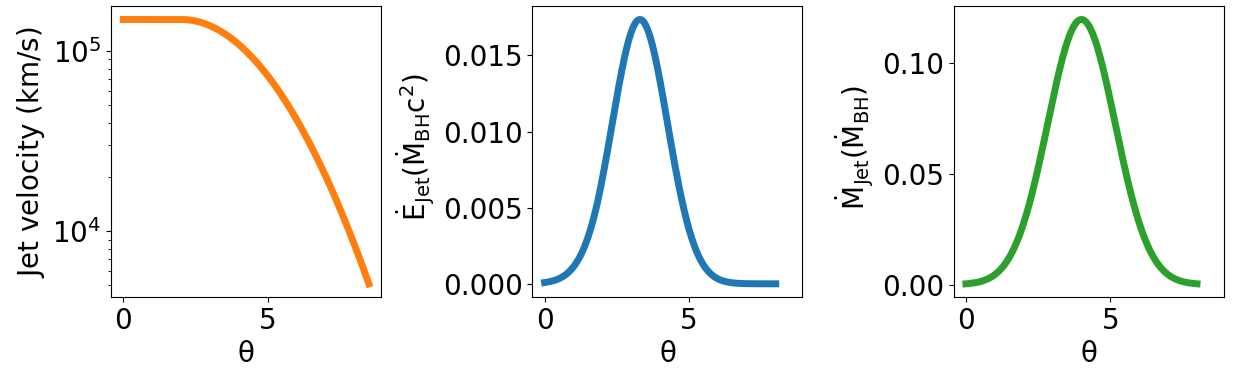}
    \caption{The jet velocity ({\it left}), kinetic power ({\it middle}) and mass flux  ({\it right}) as a function of $\theta$ adopted in our model.
\\~~~~~}
    \label{fig:jetdistr}
\end{figure*}

A two-mode AGN feedback model is adopted \citep{2018Yuan}, which distinguishes between hot and cold feedback modes based on the accretion rate of the black hole. The critical luminosity $L_{c} \sim 2\% L_\mathrm{Edd}$ separates the two modes, with the mass accretion rate corresponding to $L_{c}$ given by $\mdot_{c} \approx L_{c}/(\epsilon_\mathrm{EM,cold} C^{2})$ where $\epsilon_\mathrm{EM,cold}$ is the radiative efficiency of a cold accretion disk. The hot feedback mode (radio mode) is associated with a hot accretion flow, while the cold feedback mode (quasar mode) corresponds to high accretion rates where the accretion flow is described by the standard thin disk model \citep{1973A&A....24..337S}. The main update compared to our previous three papers of this series is the inclusion of jet. 

The parameters of the jet adopted in our model are taken from \citet{2021Yang}. They performed 3D GRMHD simulations of black hole accretion for various values of black hole spin and accretion modes  (i.e., MAD or SANE (standard and normal evolution) \citealt{2012Narayan,2015Yuan}) using the ATHENA++ code. The properties of wind and jet are obtained based on the simulation data using a ``virtual test particle approach'' initially proposed in \citet{2015Yuan}. Different from the time-averaged streamline approach widely adopted in the literature, this approach can easily discriminate turbulence and real outflow and especially provide the exact mass flux.  Observations indicate that many black holes in realistic accretion systems have spin values close to the maximum \citep{2014Reynolds}, so we assume a high black hole spin in our model. The accretion mode is poorly understood so far, and only very few sources have been investigated. One such example is M87. Detailed studies indicate that the accretion flow in M87  is MAD \citep{2021EHT,2022Yuan}. We thus assume that the accretion mode in our model is MAD.  Based on these considerations, we  have chosen the ``MAD98'' model in \citet{2021Yang} when we adopt the parameters of wind and jet, where ``MAD98'' refers to the GRMHD MAD run with black hole spin $a = 0.98$. \citet{2021Yang} initialize an equilibrium accretion torus threaded by a single poloidal magnetic loop; rapid accretion loads magnetic flux onto the hole until it saturates and the MAD state develops. One issue is that the outer boundary of their simulation domain is at $r=10^3 r_g$. This boundary is smaller than the inner boundary of our MACER simulation, which is $\sim 2.5\,pc \sim 10^4 r_g$. So we need to extrapolate the data to $10^4 r_g$. 

The first jet parameter is its mass flux $\dot M_\mathrm{jet}$. Its value is directly taken from the simulation data at $10^3 r_g$ and we assume that it remains constant out to $10^4 r_g$:
\begin{align}
    \dot M_\mathrm{jet} &= 0.35\dot M_\mathrm{BH}.
    \label{eq:mass-flux}
\end{align}

The jet mass and energy fluxes in our injection model are explicitly $\theta$-dependent, while the black hole accretion rate $\dot{M}_{\rm BH}$ is treated as a single integrated quantity with no angular dependence. 

In practice, $\dot{M}_{\rm BH}$ is first measured from the simulation, used to determine the total injected mass flux according to Equation~\ref{eq:mass-flux}, and then distributed over polar angle following the profile shown in Fig.~\ref{fig:jetdistr}.  

The angular distribution of the jet injection is adopted from the GRMHD results of ``MAD98'' \citep{2021Yang}, in which the jet mass and energy fluxes are found to be symmetric about the equatorial plane. Accordingly, jets launched from both sides of the black hole follow the same $\theta$-dependent profile in our implementation. This angular dependence is directly extrapolated from the GRMHD accretion-flow scale to the inner boundary of our simulation under the assumption of momentum conservation.

Our assumption of constant mass flux and velocity from $10^3\ r_g$ to $10^4\ r_g$ is well supported by recent studies. For jets, GRMHD simulations \citep{2019Chatterjee} demonstrate that jet mass flux is largely conserved across these scales. For winds, the mass flux is primarily determined by the underlying accretion disk structure -- specifically where the hot accretion flow transitions to a standard thin disk -- rather than conditions at the black hole horizon \citep{2015Yuan}. \citet{2020Cui} showed that across this radial range, the competing forces of gravitational deceleration versus pressure and magnetic acceleration result in approximately constant wind velocity. Gas entrainment in this region is minimal, making our simplification reasonable with negligible impact on the simulation outcomes.

The next important jet parameter is its velocity. The velocity is a function of $\theta$, as we will describe in more detail later. The peak velocity at $10^3r_g$ obtained from \citet{2021Yang} is
\begin{equation}
v_\mathrm{jet,peak} \sim 0.5c.   
\end{equation} 
We assume that it remains roughly constant from $10^3 r_g$ to $10^4 r_g$ since the jet acceleration is found to be ineffective in this region due to the change in the magnetic field morphology \citep{2021Yang}. Correspondingly, although AGN jets are known to be magnetically dominated at small scales, we neglect the electromagnetic energy of the jet in our simulations, as the jet is kinetic-energy-dominated above the radius of $10^{3}r_{g}$ \citep{2005Sikora}. The jet is injected through the inner boundary as an outflow boundary condition. This high velocity applies only to the innermost injection region, which is not shown in the current Fig.~\ref{fig:gasproperty} because our analysis focuses on the kiloparsec-scale flow where the characteristic turbulent velocities are on the order of several hundred $\mathrm{km\,s^{-1}}$. 
If one zooms in on the immediate vicinity of the injection boundary, the high-speed (\(\sim 0.5c\)) jet can be clearly observed. 
At the epoch shown in the Fig\,~\ref{fig:gasproperty}, the AGN power is relatively low, resulting in a small jet momentum flux. 
As the outflow interacts with the ambient medium, the jet decelerates substantially, and the velocity field on kiloparsec scales is therefore dominated by slower, jet-wind interaction driven motions and the associated turbulence.

The kinetic and thermal energy fluxes $\dot E_\mathrm{kin, jet}, \dot E_\mathrm{therm, jet}$ of the jet adopted in the model are as follows:
\begin{align}
       \dot E_\mathrm{kin, jet} &= \frac{1}{2} \dot M_\mathrm{jet} \bm{v}_\mathrm{jet}^{2}, \\
    \dot E_\mathrm{therm, jet} &= \frac{1}{3} \dot E_\mathrm{kin, jet},
\end{align}

The jet is injected through the inner boundary as an outflow boundary condition.   
At the initial moment, the jet is highly supersonic, so all characteristic waves at the boundary point inward toward the computational domain. 
Under these conditions, the fluxes of momentum and energy are completely determined by the jet properties at the boundary. 
As a result, both the injected momentum and energy are exactly conserved in the numerical scheme. 
The momentum flux is injected within the prescribed angular sector, and the corresponding kinetic and internal energy fluxes are computed consistently from the injected momentum flux, ensuring full numerical self-consistency.

Another important parameter is the half-opening angle of the jet, $\theta_\mathrm{j} \equiv \arctan(R/Z)$, where $R$ and $Z$ are the half-radius and the length of the jet, respectively. \citet{2021Yang} find that the half-radius and the length of the jet follows a power-law profile $R=1.01\ Z^{0.8}$. Therefore, we calculate the half-opening angle of the jet using this power-law profile at the inner boundary ($\sim10^4 r_g$) of our simulation domain and find:
\begin{equation}
\theta_\mathrm{j}\approx  7.5^\circ.
\end{equation}
This value is roughly consistent with the observational result presented in \citet{2016Mertens}.

The velocity distribution along the spherical $\theta$ coordinate $\bm{v}_\mathrm{jet}(\theta)$ between $\theta=0$ and $\theta=\theta_{\mathrm{j}}$ is also important since it affects the interaction between the jet and wind and the subsequent production of turbulence, as we will show later. Following the simulation results presented in \citet{2021Yang}, we set the velocity distribution to be a Gaussian function beyond $\theta\sim 2 ^\circ$ (because there is no data within this angle due to the singularity of spherical coordinates), with a peak velocity $v_\mathrm{jet,peak} \sim 0.5c$ at $\theta_\mathrm{jet, peak} \sim 2^\circ$ and an angular dispersion $\sigma \sim 1.17^\circ$. The left plot of Fig.~\ref{fig:jetdistr} shows the jet velocity as a function of $\theta$. This distribution, combined with the $\theta$-dependence of the gas density in the jet we have obtained from \citet{2021Yang}, gives us the kinetic power and mass flux as a function of $\theta$. The results are shown in the middle and right plots of this figure, respectively.

The jet model simulated in \citet{2021Yang} is the Blandford-Znajek (BZ) jet. While this jet model is influential and has been confirmed by GRMHD numerical simulations \citep{2014Yuan}, it only deals with the dynamics of jets. One may wonder whether the jet can explain various observational results, including the spectral energy distribution and many morphological results.  This question has been addressed in \citet{2024Yang}. By assuming the electrons in the jet are accelerated by magnetic reconnection in the jet, they have calculated the spatial and energy distribution of nonthermal electrons in the jet. They then calculated the radiative transfer and obtained much observational information such as the elongated structure of the jet, brightness distribution of the jet (e.g., the limb-brightening feature), and jet width as a function of distance. They then compared these results with observations and found good consistency. They also calculated the predicted image by another jet model, i.e., the Blandford-Payne jet model, and found that the predicted results are not consistent with observations; thus, this model is ruled out. Their results show that the BZ jet we have adopted should correspond to the observed ``real jet.''

We do not consider jet precession in our simulations. This is motivated by the observations which show that most AGN jets don't have precession. Only less than 10\% of radio sources exhibit S- or Z-shaped morphologies, suggesting that the majority of jets do not undergo precession \citep{2011Proctor}.

\subsection{Summary of the simulation suit}

Our simulation suite consists of three representative runs, {\tt FullFeedback}, {\tt WindOnly}, and {\tt JetOnly}, whose names are self-explanatory: the {\tt FullFeedback} run includes both jet and wind feedback, and the others only include wind or jet feedback, respectively. 
Our {\tt WindOnly} setup is close to the fiducial simulation setup of \citep{2018Yuan}, which is designed to compare with previous work and study the effects of jet-wind interactions on the evolution of massive elliptical galaxies.
We assume that the jet only exists in the hot mode when the AGN accretion rate is low \citep{2014Yuan}, so jet feedback is turned on in {\tt FullFeedback} and {\tt JetOnly} only when the AGN enters the hot mode\footnote{Sometimes jets likely also exist in the cold mode, when the AGN accretion rate is high. This is evidenced by the fact that about 10\% of the quasars are radio-loud. However, it is still an unsolved problem how jets are formed from an accretion flow with a high accretion rate. }. In addition, we note that in the {\tt JetOnly} run, only in the hot mode (when the jet is present) is the wind turned off, while we do \emph{not} turn off the wind in the cold mode (when the jet is absent), since a cold mode with neither wind nor jet feedback is not unphysical.

Following \citet{2018Yuan}, all simulations start from a stellar population age at $t = 2\,\mathrm{Gyr}$ and are evolved for $t = 12\,\mathrm{Gyr}$.

\section{Results and Disscussion}
\label{sec:results}

\begin{figure}
    \centering
    \includegraphics[width=0.45\textwidth]{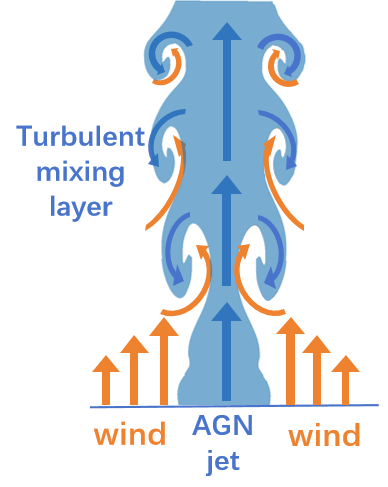}
    \caption{A schematic figure showing the production of turbulence due to the KH instability driven by the velocity shear between jet and wind. In this way, the kinetic energy of the jet is efficiently transformed into turbulence and finally dissipated into the ISM.}
    \label{fig:cartoon}
\end{figure}

Before we introduce our simulation results, we first illustrate in the following section an important physical process that we identify in our model, that is, the nonlinear interaction between jet and wind and the resulting turbulence. We find that this process efficiently transfers the energy of the jet to the ISM, thus significantly affecting the SFR.

\subsection{AGN jets-wind interaction}
\label{sec:coupling}

Fig.~\ref{fig:cartoon} is a schematic figure showing the interaction between jet and wind and the production of turbulence. The velocity of AGN jets is approximately an order of magnitude higher than that of the wind.  This shear in velocity across the boundary layer manifests as a velocity gradient, forming a so-called tangential discontinuity surface in viscous fluids. Consequently, Kelvin-Helmholtz (KH) instability develops. The development of the instability results in the production of turbulence \citep{1987Landau,2000Pope}. In this process, the kinetic energy of the bulk motion of the jet will first be transformed into turbulent energy, which will be further dissipated and converted into thermal energy through a cascade process. In short, the nonlinear coupling between the jet and wind results in a much higher energy deposit efficiency compared to their linear sum, as we will discuss in Section \ref{sec:dissipationefficiency}.

In the {\tt JetOnly} case, the KH instability also exists due to the velocity shear between jet and ISM. But whether the instability can grow also depends on the comparison between the instability growth timescale and the dynamical timescale. We have calculated the ratio of the KH instability growth timescale and the dynamical timescale in {\tt FullFeedback} and {\tt JetOnly} models. We find that the ratio in the former is several times smaller than the latter. This result suggests that the turbulence in the case of {\tt FullFeedback} will be much stronger than that in the case of {\tt JetOnly}. This question will be discussed in detail in a companion paper \citep{2026Guo}. In the present paper, we focus on the effects of this coupling on the evolution of an elliptical galaxy. In another companion paper, we show that this mechanism successfully solves the long-standing cooling flow problem in galaxy clusters (He et al. 2025).

\subsection{Spatial distribution of gas properties}

\begin{figure*}
    \centering
    \includegraphics[width=1\textwidth]{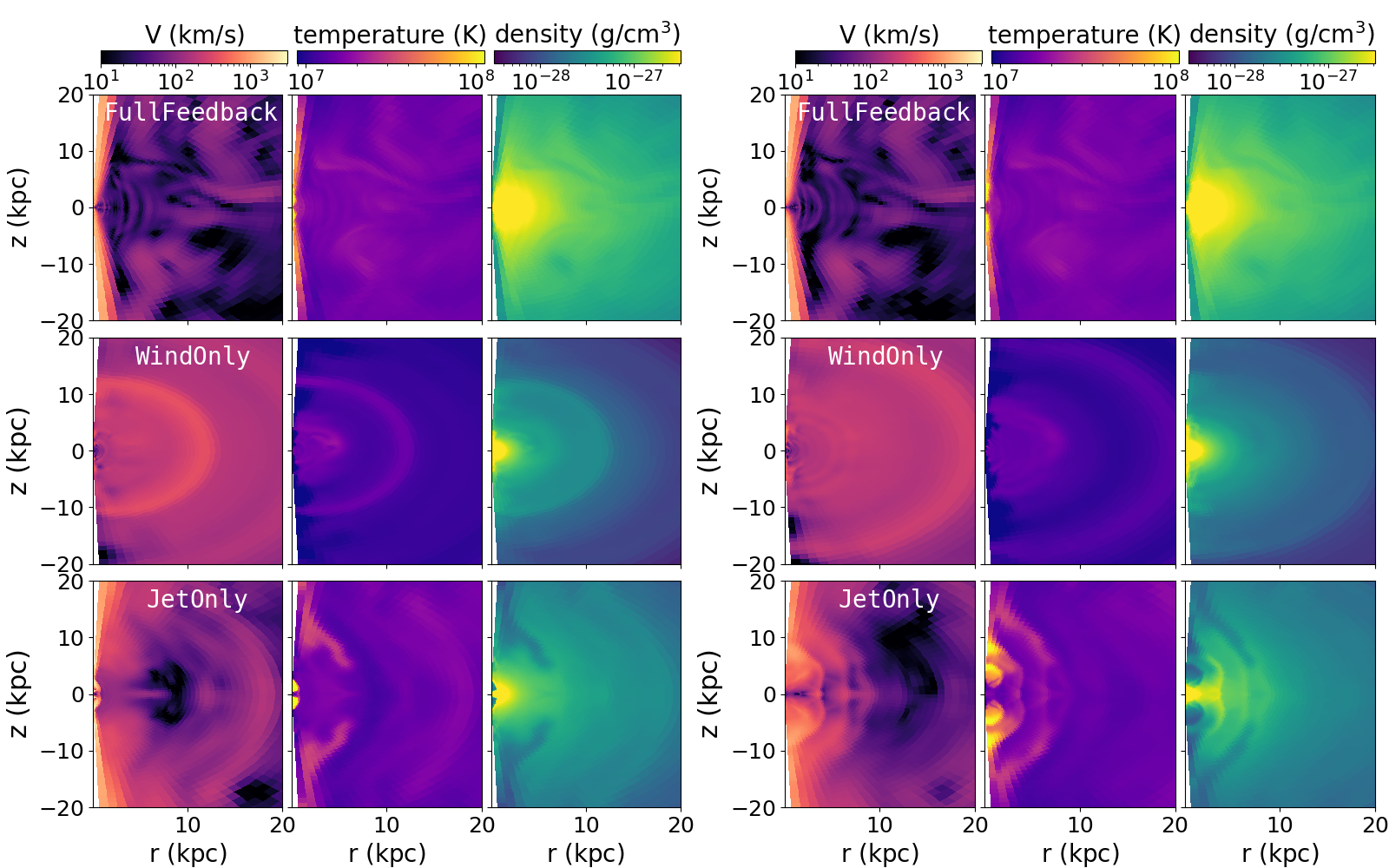}
    \caption{
        The zoomed-in ($r\lesssim 20\,\mathrm{kpc}$) spatial distribution of gas properties in the runs {\tt FullFeedback} (top), {\tt WindOnly} (middle) and {\tt JetOnly} (bottom) at $t\sim 7.58\,\rm{Gyr}$ (left) and $t\sim 7.6\,\rm{Gyr}$ (right)  when the AGN is in the hot mode: (from left to right) the velocity magnitude (km/s), temperature (K), and gas density (g/cm\(^3\)). All colorbars are plotted in logarithmic scale. Compared to the {\tt WindOnly} run, strong anisotropy in gas properties arises in the {\tt FullFeedback} and {\tt JetOnly} runs.
         \\~~~~~
        } 
  \label{fig:gasproperty}
\end{figure*}

Fig.~\ref{fig:gasproperty} shows  the spatial distribution of velocity, temperature, and gas density at $t\sim 7.58$ and $t\sim 7.6\,\rm{Gyr}$ from three simulations: {\tt FullFeedback} (top), {\tt WindOnly} (middle), and {\tt JetOnly} (bottom). The snapshots are deliberately chosen when the AGN is in the hot mode and the jet exists. In all simulations, the outflowing gas driven by wind and jet propagates mainly along the polar direction, while the infalling cool, high-density gas is primarily along the equatorial plane. As expected, strong anisotropy in gas properties arises in the {\tt FullFeedback} and {\tt JetOnly} runs. In particular, the jet regions near the vertical axis are dominated by hot (up to $10^{10}\,\mathrm{K}$), low-density, and high-velocity (up to a few $10^3\,\mathrm{km/s}$) gas. From the perspective of the definition of entropy $K\sim Tn^{-2/3}$, these gases exhibit high entropy, suggesting a significant amount of local jet heating. In contrast, the gas properties are more isotropic in the {\tt WindOnly} run.

The comparison of the morphology of the jet in  {\tt FullFeedback} and {\tt JetOnly} reveals important clues about jet collimation. In the {\tt FullFeedback} run, the jet remains well confined by the surrounding wind, maintaining its initial half-opening angle and extending up to $\sim 10^4 \,\mathrm{pc}$ along the $z$-axis without significant side expansion. However, in the {\tt JetOnly} run, the jet can only be marginally confined by the ambient gas pressure within a distance of $\sim 10^2\,\mathrm{pc}$. Beyond this distance, it significantly expands. This result of jet collimation by wind is consistent with the  analytical analysis in \citet{2019Park}. 

The second column depicts temperature variations. The {\tt FullFeedback} model shows hotter regions compared to the other two cases, especially at larger radii. This indicates that the combination of jets and winds is more efficient at heating the gas, which could prevent cooling and suppress star formation. The {\tt WindOnly} and {\tt JetOnly} models exhibit cooler gas in comparison, with {\tt JetOnly} showing a slight temperature increase in the central regions due to the localized impact of jet feedback. The density panels reveal important differences in gas distribution among the three models. The {\tt FullFeedback} model shows lower central densities, implying that the feedback is effectively pushing gas away from the galaxy center.  The {\tt WindOnly} and {\tt JetOnly} models, on the other hand, show higher central densities, particularly in {\tt JetOnly}, where the jets are less efficient at clearing out the gas. This reinforces the idea that a combination of winds and jets in the {\tt FullFeedback} case leads to more efficient energy dissipation and gas removal, which could have significant implications for star formation suppression, as we will discuss in the next section.

As a quick recap, for jet versus nonjet runs, the inclusion of the AGN jet feedback can result in significant changes in the thermal and kinematic properties of the gas, which is not surprising; however, a more interesting hint is that the effects of the jet feedback strongly depend on whether the jet is surrounded by the wind or not. This point becomes more evident in the following sections.

\subsection{Star formation}

\begin{figure*}
    \centering
    \includegraphics[width=0.95\textwidth]{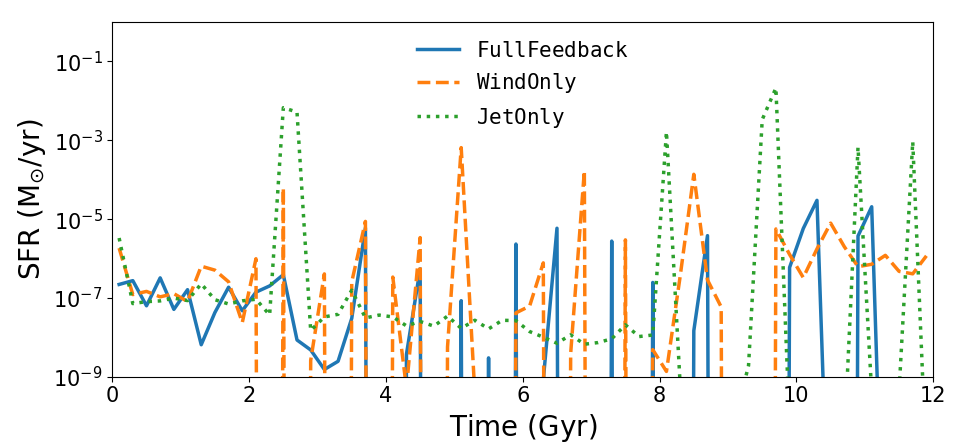}
    \caption{The evolution of SFR over 12 Gyr for three simulations. The {\tt JetOnly} simulation exhibits the highest time-averaged SFR at approximately $10^{-1} \mathrm{M}_\odot\,\mathrm{yr}^{-1}$. The {\tt WindOnly} simulation shows a time-averaged SFR one order of magnitude lower at $10^{-2} \mathrm{M}_\odot\,\mathrm{yr}^{-1}$, while the {\tt FullFeedback} simulation maintains the lowest time-averaged SFR at $10^{-3} \mathrm{M}_\odot\,\mathrm{yr}^{-1}$. Note: The time-averaged SFR is calculated as the total mass of newly formed stars divided by the total simulation time. Some SFR peaks may not be fully resolved due to temporal sampling, potentially leading to apparently low SFR values as shown in this figure.
 \\~~~~~}
    \label{fig:interSFR}
\end{figure*}

\begin{figure*}
    \centering
    \includegraphics[width=0.95\textwidth]{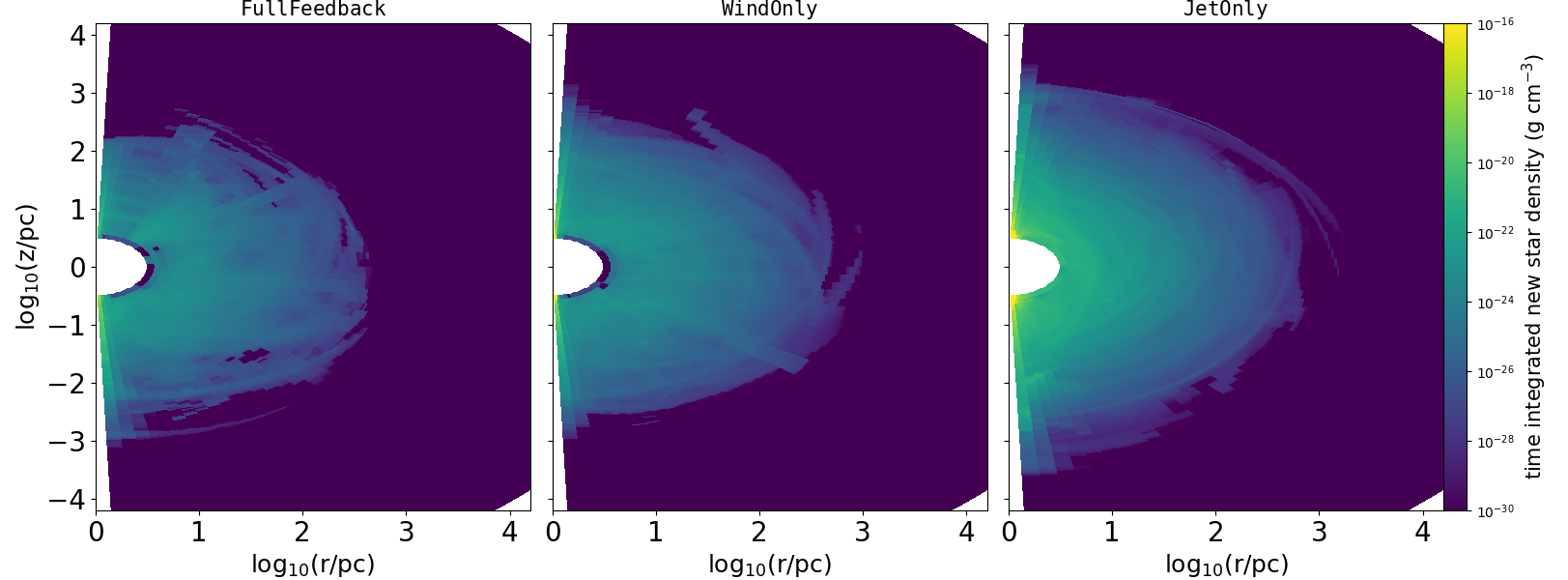}
    \caption{The distribution of time-integrated mass density of newly born stars at the end of the run for {\tt FullFeedback}, {\tt WindOnly},  {\tt JetOnly} simulations, respectively. The {\tt JetOnly} simulation shows a much higher newly formed star density and a more extended star formation region, because without winds, the energy deposition efficiency of jet alone is very low.
 \\~~~~~}
    \label{fig:starformation_density}
\end{figure*}

Fig.~\ref{fig:interSFR} illustrates the temporal evolution of SFR for the three simulations. Fig.~\ref{fig:starformation_density} displays the spatial distribution of time-integrated newborn star mass density for all three simulations. 

We have calculated the time-averaged SFR, defined as the total mass of newly formed stars divided by the total simulation time. We find that the {\tt JetOnly} simulation exhibits the highest time-averaged SFR of $\sim 10^{-1} \mathrm{M}_\odot\,\mathrm{yr}^{-1}$, while the {\tt WindOnly} simulation shows a time-averaged SFR one order of magnitude lower, at $10^{-2} \mathrm{M}_\odot\,\mathrm{yr}^{-1}$. The {\tt FullFeedback} simulation maintains the lowest time-averaged SFR, which is $10^{-3} \mathrm{M}_\odot\,\mathrm{yr}^{-1}$. 

The 1 order-of-magnitude lower SFR than {\tt WindOnly} in the {\tt FullFeedback} model and two-order-of-magnitude lower SFR than {\tt JetOnly} indicates that the {\tt FullFeedback} simulation does not simply combine the effects of jet and wind; instead, it demonstrates that the wind-jet coupling leads to a more effective suppression of star formation than the ``linear sum'' of the jet and wind, as we discuss in \S\ref{sec:coupling}. 

The SFR in the {\tt JetOnly} simulation is the highest among the three models. This can be attributed to the inefficient heating effect of AGN jets. This is especially the case given that the typical AGN power in this model is higher than the other two models, as will be discussed in Section \ref{sec:agnlightcurve}.  Without AGN hot winds, jets fail to effectively deposit their energy into the ISM and suppress star formation. This result aligns with previous studies using nonprecessing jet simulations. The jet alone fails to maintain the quiescent state of early-type galaxies — a behavior that contradicts observations. This is why some additional mechanisms such as the ``dentist drill'' \citep{2006Vernaleo} effect have been proposed in the literature to overcome the problem.

\begin{figure*}
    \centering
    \includegraphics[width=1.0\textwidth]{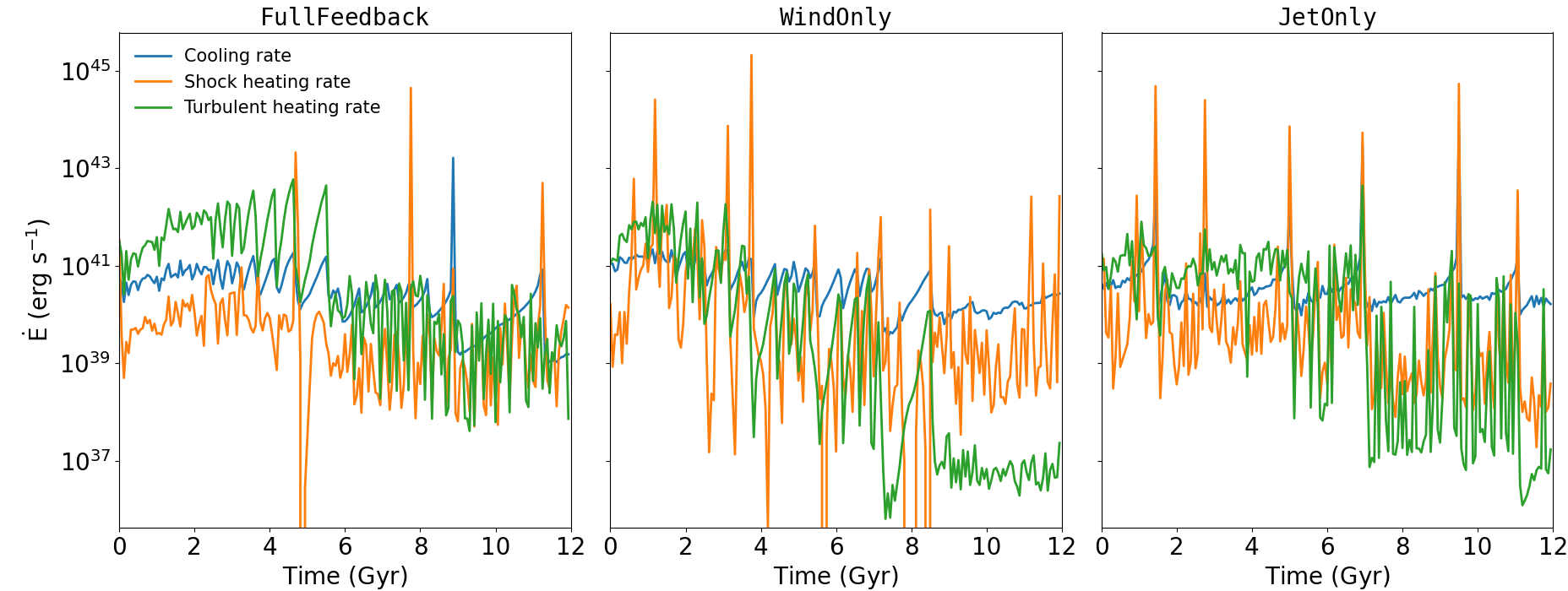}
        \caption{Temporal evolution of the cooling and heating rates within a spherical region of radius $R<35\rm{kpc}$ in {\tt FullFeedback}, {\tt WindOnly}, and {\tt JetOnly} runs. The blue line shows the cooling rate due to radiative processes, the orange line for the estimated shock heating rate, and the green line for the turbulent heating rate. Turbulent heating is  more efficient than shock heating in {\tt FullFeedback} simulation.}
        \label{fig:h/c}
\end{figure*}

On the other hand, the time-averaged SFR in the {\tt WindOnly} model is smaller than that in the {\tt JetOnly} model by 1 order of magnitude. As we will show in Section \ref{sec:agnlightcurve}, the AGN power in this model is lower than that in the {\tt JetOnly} model. In addition,  the power of jets is about 10 times higher than the wind for a rapidly spinning black hole \citep{2021Yang}. Therefore, wind is more efficient in suppressing star formation than jets. There are two reasons for this result. The first reason is that the energy deposition efficiency of jets is lower than wind. This is because the opening angle of jets is much smaller than wind. The second reason is that the efficiency of star formation suppression depends not only on the energy of jets and wind but also on their momentum. \citet{2021Yang} show that the momentum flux of wind is larger than jets. 

We also note a concentration of newly formed stars near the polar axis, associated with the so-called funnel accretion \citep{Proga2003,Ressler2018,Yao2020}, which arises from the inflow of low angular momentum gas. While our numerical setup (polar coordinates with reflective boundary conditions at the axis) may influence the exact distribution, the significant differences in star formation patterns between our three simulations, which share identical numerical geometry, indicate that this feature primarily reflects the underlying physical mechanisms of the anisotropic AGN feedback and the fallback of cooled low angular momentum gas.

\subsection{Heating/cooling rates and energy budget}
\label{sec:H/C}

The heating and cooling processes occurring in the galaxy include the radiative heating and cooling, the turbulent cascade, and heating by shock waves. 
The shocks and turbulence are driven by the interaction between the wind, jet, and ISM. 
In this section, we focus mainly on the latter two processes, i.e., the heating caused by the dissipation of the turbulent cascade and shock waves.

\subsubsection{The turbulent heating rate}

To estimate the energy dissipation rate associated with turbulent motions, we perform a spectral analysis of velocity fluctuations with respect to the radial motion of the gas. 
Only a fraction of such fluctuations can be attributed to turbulence; therefore, they can only be used to infer an upper limit to the turbulent heating rate. Previous work shows that strong shocks also introduce turbulence that can efficiently transport the shock energy to the ambient gas in the post shock region \citep{2016Ji}. For these reasons, the ``turbulent heating rate'' estimated below represents an upper limit of the heating rate provided by both a turbulent cascade from large scales down to the grid scale and gas mixing occurring on large scales.

Our direct numerical simulation involves solving the Navier-Stokes equations without relying on Reynolds-averaged or large-eddy simulation models.
We focus on velocity fluctuations relative to a background that may have net radial velocity (due to AGN activities, the gas velocity field is dominated by radial velocity). We define the velocity fluctuations as
\begin{equation}
 \delta \vec{v}(x) = \vec{v}(x)-\vec{v}_{\rm rad}(x),
\end{equation}
where $\vec{v}_{\rm rad}(x)$ is the radial velocity at radius $x$. Based on Kolmogorov's first similarity hypothesis \citep{2000Pope}, in turbulent flows with sufficiently high Reynolds numbers, eddies possessing energy on the order of $\delta \vec{v}^2(k)$ with a turnover timescale defined by $\tau(k) = {\delta \vec{v}(k)}\cdot k$ effectively dissipate kinetic energy at a rate given by nearly equivalent to the rate of energy transfer from larger scales.

According to Kolmogorov's second similarity hypothesis, the statistical properties of motions at scale $k$ are determined by the turbulent dissipation rate $\epsilon$. Finally, we estimate the kinetic energy dissipation rate as:
\begin{equation}\label{eq:turb}
 \dot{E}_{\rm kin,turb} = \frac{1}{2}\bar{\rho}[\sigma^2(k)]^{3/2}k,
\end{equation}
where $\bar{\rho}$ is the average density in the region where the power spectrum is computed, $\sigma^2(k)$ is the velocity
dispersion of the velocity fluctuations with respect to the radial velocity field estimated at the effective driving scale $k$. Equation~\ref{eq:turb} is derived by assuming that the kinetic energy $E_{\rm kin,turb}\sim 0.5\bar{\rho}\sigma^2$ associated with turbulent eddies of size $L\sim1/k$ will cascade down to eddies of smaller size in approximately one eddy turnover time.

\begin{figure*}
    \centering
    \includegraphics[width=0.99\linewidth]{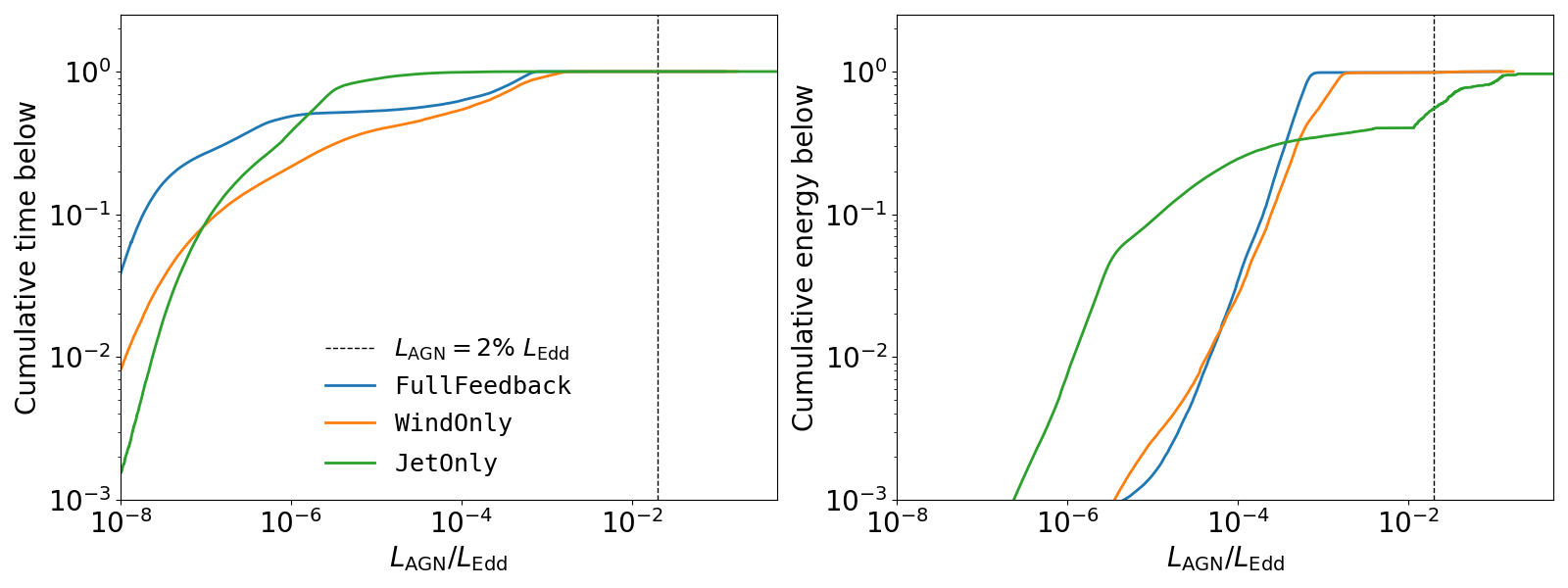}
    \caption{Percentage of the total simulation time spent ({\it left}) and cumulative energy emitted ({\it right}) below the given values of AGN Eddington ratios. In the {\tt JetOnly} model,  cold  mode (i.e., above $2\% L_{\rm Edd}$) occupies very little temporal residence but accounts for approximately 50\% of the integrated AGN energy. }
    \label{fig:duty_cycle}
\end{figure*}

\begin{figure*}
    \centering
    \includegraphics[height=0.55\textheight]{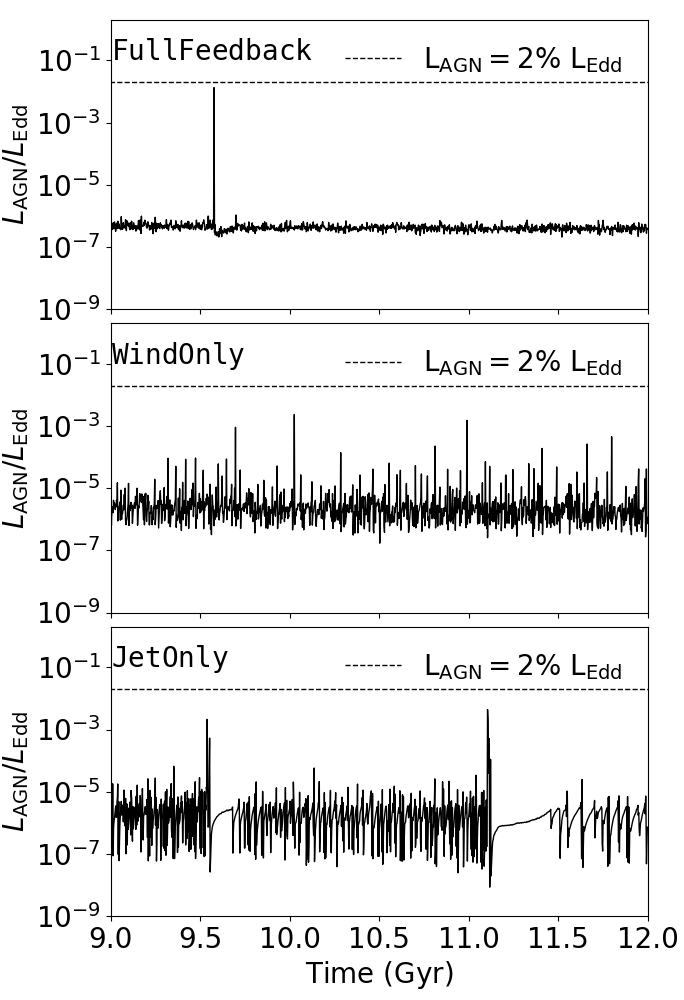}\hspace{0.02\textwidth}
    \includegraphics[height=0.55\textheight]{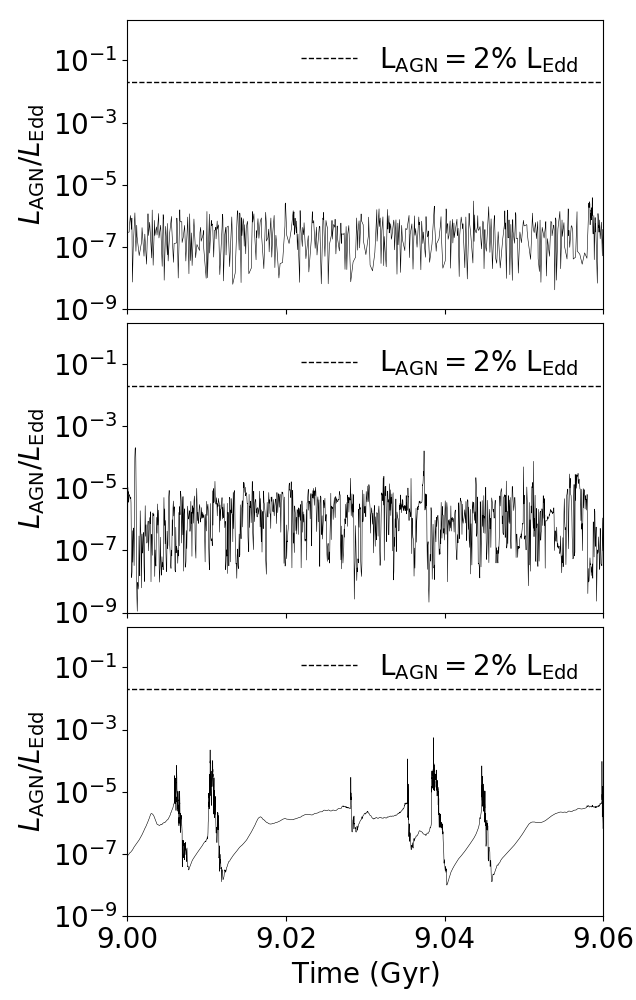}
    \caption{Temporal evolution of AGN luminosity in the {\tt FullFeedback} (upper), {\tt WindOnly} (middle), and {\tt JetOnly} (bottom) simulations during the late evolutionary phase. The right panels show the zoomed-in part of the left panels. The dashed line indicates the critical luminosity $ L_c \equiv 0.02L_{\rm Edd}$. }
    \label{fig:Lum}
\end{figure*}

\subsubsection{The shock heating rate}

We adopt a linear approximation method of shock heating which uses theoretical arguments similar to those used by \citet{1987Landau}, \citet{2016Yanga,2016Yangb}, and \citet{2019Martizzi}. Shocks are identified in our simulations by measuring entropy, pressure, and density jumps. The pressure jump across a shock can be expressed as:
\begin{equation}\label{eq:dpoverp}
\frac{\Delta P}{P} = \frac{P_2-P_1}{P_1} = \frac{2\gamma}{\gamma+1}y,
\end{equation}
where $P_1$ and $P_2$ are the pre shock and post shock pressures, respectively, and $y$ is a dimensionless parameter related to the shock Mach number $M_{\rm s}$:
\begin{equation}\label{eq:mach}
y = \frac{\rho_1v_{\rm s}^2}{\gamma P_1} -1 = M_{\rm s}^2 - 1,
\end{equation}
where $\rho_1$ is the pre shock density. The density jump across the shock is given by:
\begin{equation}\label{eq:rhojump}
\delta = \frac{\rho_2}{\rho_1} = \frac{(\gamma+1)(y+1)}{2+(\gamma-1)(y+1)}.
\end{equation}
Finally, the entropy jump across a shock is given by:
\begin{equation}
ds \approx \frac {2\gamma k_{\rm B}}{3(\gamma+1)^2\mu m_{\rm H}}y^3.
\end{equation}
We measure density, pressure, and entropy jumps in all cells in the computational volume by computing differences between post shock and pre shock conditions. A cell is flagged as shock heated only when $\nabla\cdot v<0$, $ds > 0$, and $\Delta P/P > 0.1$. The first condition ensures that we are measuring density jumps that are representative of shocked regions, i.e., where the density jump is similar to that given by the shock jump condition of equation~\ref{eq:rhojump}. The second condition ensures that there is local heating, which is associated with an entropy jump. The third condition ensures that we detect shocks with a Mach number above a given threshold.
Once $\Delta P/P$ is measured, $y$ is calculated from equation~\ref{eq:dpoverp} and associated with a Mach number using equation~\ref{eq:mach}. 

Finally, once a cell has been flagged as shock heated, the shock heating rate is estimated as:
\begin{equation}\label{eq:shock}
\dot{E}_{\rm shock} = \frac{\rho_1T_1ds}{\Delta t},
\end{equation}
\begin{equation}
\Delta t = \frac{v_s}{\Delta r},
\end{equation}
where $T_1$ is the pre shock temperature, $v_s = M_s \cdot c_s$ is the shock velocity, $c_s$ is the local sound speed, and $\Delta r$ is the distance of shock propagation, which is treated as equal to the grid size.

\subsubsection{Comparison between turbulent dissipation and shock heating}
Figure~\ref{fig:h/c} illustrates the temporal evolution of cooling and heating rates within a spherical region of radius $R<35\,\mathrm{kpc}$ for the three models.  The radiative cooling rate remains at approximately $\mathrm{10^{42}\ erg\,s^{-1}}$. We find that the total heating rate, including turbulent and shock heating, roughly correlates with AGN luminosity, as turbulent motion and shock waves are introduced by AGN activities through jets and winds. 

We have calculated the time-integrated heating and cooling energy and found that the cooling and heating are roughly balanced in the three models. More specifically, in {\tt FullFeedback}, turbulent heating surpasses the shock heating and is comparable to the gas radiative cooling. 
But in {\tt WindOnly} and {\tt JetOnly} models, the turbulent dissipation is weaker than the shock heating. Defining $\beta_{E_T/E_S}$ as the ratio of turbulent heating to shock heating, we find that its values are 4.9, 0.52, and 0.29 for {\tt FullFeedback}, {\tt WindOnly}, and {\tt JetOnly}, respectively. 
We think that the reason for the above differences among the three models is that, in {\tt JetWind} model, the turbulence produced by KH instability driven by the jet-wind shear is the strongest \citep{2026Guo}. 

\subsubsection{AGN energy dissipation efficiency}
\label{sec:dissipationefficiency}

We define an AGN dissipation efficiency as the fraction of the injected AGN kinetic energy that is dissipated via turbulence and shock to heat the ISM:
\begin{equation}
\eta_{\rm dissipation} = \frac{{E}_{\rm{turbulent\ heating}}+{E}_{\rm{shock\ heating}}}{{E}_{\rm AGN\ kinetic}},
\end{equation}
where ${E}_{\rm{turbulent\ heating}}$ and ${E}_{\rm{shock\ heating}}$ represent the time-integrated dissipated energy through turbulent and shock heating mechanisms (calculated using Equations~\ref{eq:turb} and \ref{eq:shock}, respectively), and ${E}_{\rm AGN\ kinetic}$ denotes the total kinetic energy input from AGN (including both jets and winds) over the 12 Gyr simulation period. Our calculations reveal that the {\tt FullFeedback} model achieves an efficiency of 0.64, substantially exceeding the values obtained in single-mode feedback models: 0.48 for {\tt WindOnly} and 0.26 for {\tt JetOnly}. This result  again provides a compelling evidence that the non-linear coupling between AGN jets and winds serves as a highly effective mechanism for energy conversion and dissipation in feedback processes.

\subsection{AGN light curve and duty-cycle}
\label{sec:agnlightcurve}

Fig.~\ref{fig:duty_cycle} shows the percentage of the total simulation time spent (left) and cumulative energy emitted (right) below the given values of AGN Eddington ratios for the three models. Significant variations in the AGN duty cycles across different models are evident. For the three models, during most of the evolution, the AGN stays in the hot mode, consistent with observational data from the SDSS (\citealt{2009Kauffmann}). 
In the {\tt JetOnly} model, although the cold mode has a short lifetime, it nevertheless contributes approximately 50\% of the integrated energy emitted by the AGN.

Figure~\ref{fig:Lum} shows the temporal evolution of the AGN luminosity in the three models. From the figure, we can see that the AGN luminosity in the {\tt JetOnly} model is the highest, then the {\tt WindOnly} model, and the luminosity in the {\tt FullFeedback} model is the lowest. Such a sequence is because of the AGN energy dissipation efficiencies of the three models, as we have discussed in section \ref{sec:dissipationefficiency}. When the efficiency is high, the heating will be efficient; thus, the gas will in general stay in a high-temperature-low-density state, which results in a low luminosity. 

\section{Conclusions}

\label{sec:diss}

In the first three works of our series of works on the effects of AGN feedback in the evolution of an elliptical galaxy, for the AGN outputs, we only considered radiation and wind, but neglected jet. As the fourth paper of this series, the main aim of the present work is to include jets and investigate the effect of jets in AGN feedback. For this aim, we have simulated three models, namely {\tt FullFeedback}, {\tt WindOnly}, and {\tt JetOnly}. The main results are summarized below:
\begin{itemize}
\item 
Most previous works studying jet feedback treat the jet parameters as free. These parameters include the mass flux, velocity and its distribution in the cross section of the jet, and the opening angle. The values of the jet parameters in those works are unfortunately often unphysical. Different from these works, in our model, the values of the parameters of AGN jets are adopted from the GRMHD numerical simulation of black hole accretion and jet formation \citep{2021Yang}\footnote{Note that the parameters of winds in {\it MACER} are also taken from the accretion flow-scale GRMHD numerical simulation work.}. We find that the correct setting of the jet and wind parameters is important because the wind-jet shear results in the development of turbulence, as we describe below. 

\item We have investigated the effects of the jet and wind on the SFR. The time evolution of the SFR for the three models is shown in Fig.~\ref{fig:interSFR}. The time-averaged values of the SFR are $10^{-1}$, $10^{-2}$, and $10^{-3} \mathrm{M}_\odot\,\mathrm{yr}^{-1}$ in the {\tt JetOnly}, {\tt WindOnly}, and {\tt FullFeedback} models, respectively. So   wind feedback is more efficient than the jet, which is because the opening angle of the wind is much larger than the jet and the momentum flux of the wind is larger than the jet \citep{2021Yang}. The very low SFR in the {\tt FullFeedback} model indicates that there exists nonlinear coupling between the wind and the jet, whose feedback effect is much stronger than the linear sum of the jet and wind. This coupling is the KH instability driven by the wind-jet shear. The instability drives turbulence, whose dissipation effectively heats the gas and suppresses star formation. 

\item The heating rate by turbulent dissipation and shocks  and their comparison with the radiative cooling are shown in Fig.~\ref{fig:h/c}. In the {\tt FullFeedback} simulation, the turbulent heating rate is larger than the shock heating rate, and their sum is comparable to the gas radiative cooling rate. But in the {\tt WindOnly} and {\tt JetOnly} simulations the shock heating rate is stronger than the turbulent dissipation, although the sum of the shock heating and turbulent dissipation is also comparable to the cooling rate. The reason for the difference of the dominant heating mechanism is that the turbulence produced by the KH instability driven by the jet-wind shear is much stronger than the turbulence produced in the other two models \citep{2026Guo}. 

\item We have calculated the AGN energy dissipation efficiency, which is defined as the fraction of the injected AGN kinetic energy that is dissipated via turbulent dissipation and shock heating. The values of the efficiency for the three models are 0.64 ({\tt FullFeedback}), 0.48 ({\tt WindOnly}), and 0.26 ({\tt JetOnly}), respectively. The highest efficiency in {\tt FullFeedback} is also because of its strongest turbulence driven by wind-jet coupling. 
 
\item In all three models, the predicted duty cycle is consistent with observations. {\tt FullFeedback} has the lowest typical AGN luminosity, which is again because the AGN energy dissipation efficiency in this model is the highest.

\end{itemize}

\begin{acknowledgments}

\emph{Acknowledgments} We thank the anonymous referee for constructive comments that improved the paper. We thank H. W. Chen, L. C. Ho, and Z. Qu for stimulating discussions. The authors are supported by the National Key R\&D Program of China No. 2023YFB3002502, the NSF of China (grants 12133008, 12192220, 12192223, and 12361161601), and the China Manned Space Program (grants CMS-CSST-2025-A08 and CMS-CSST-2025-A10). This work was performed in part at the Aspen Center for Physics, which is supported by National Science Foundation grant PHY-2210452. Numerical calculations were run on the CFFF platform of Fudan University, the supercomputing system in the Supercomputing Center of Wuhan University, and the High Performance Computing Resource in the Core Facility for Advanced Research Computing at Shanghai Astronomical Observatory. 
\end{acknowledgments}

\software{{\small Matplotlib} \citep{hunter2007matplotlib},
          {\small NumPy} \citep{van2011numpy}, 
          {\small SciPy} \citep{oliphant2007python}, 
          {\small yt} \citep{Turk2010,turk2024introducing}
          }

\bibliographystyle{aasjournal}
\bibliography{main}

\end{document}